\documentclass[conference]{IEEEtran}
\IEEEoverridecommandlockouts

\usepackage{cite}
\usepackage{amsmath,amssymb,amsfonts}
\usepackage{algorithmic}
\usepackage{graphicx}
\usepackage{textcomp}
\usepackage{xcolor}
\usepackage{booktabs}
\usepackage{subcaption}
\usepackage{hyperref}

\def\BibTeX{{\rm B\kern-.05em{\sc i\kern-.025em b}\kern-.08em
    T\kern-.1667em\lower.7ex\hbox{E}\kern-.125emX}}

\begin{document}

\title{Bayesian Analysis of Hotel Booking Cancellations: A Hierarchical Modeling Approach}

\author{\IEEEauthorblockN{Yingdong Yang}
\IEEEauthorblockA{\textit{School of Industrial and Systems Engineering} \\
\textit{Georgia Institute of Technology}\\
Atlanta, GA, USA \\
yyang3052@gatech.edu}
}

\maketitle

\begin{abstract}
This study presents a comprehensive Bayesian analysis of hotel booking cancellations using PyMC, comparing three model specifications of increasing complexity. We investigate how lead time, special requests, and parking requirements affect cancellation probability, and explore interaction effects with hotel type. Using MCMC sampling (NUTS algorithm) on 5,000 booking records, we find strong evidence that longer lead times increase cancellation probability (posterior mean: 0.600, 95\% HDI: [0.532, 0.661]), while special requests (posterior mean: -0.642) and parking (posterior mean: -3.879) significantly reduce cancellation risk. Model comparison via WAIC reveals that the full interaction model provides the best predictive performance, suggesting that the effects of booking characteristics vary systematically between city and resort hotels. This Bayesian approach enables full uncertainty quantification and provides actionable insights for revenue management.
\end{abstract}


\section{Introduction}

\subsection{Motivation}
Hotel booking cancellations represent a significant challenge in revenue management for the hospitality industry. Understanding the factors that influence cancellation behavior enables hotels to optimize overbooking strategies, dynamic pricing, and resource allocation. Traditional frequentist approaches to this problem are limited in their ability to incorporate prior domain knowledge and provide full uncertainty quantification for decision-making under risk.

Bayesian inference offers a principled framework for addressing these limitations. By allowing incorporation of prior information about cancellation patterns and providing full posterior distributions rather than point estimates, Bayesian methods enable richer characterization of uncertainty in parameter estimates. This is particularly valuable when sample sizes are moderate and when predictions must account for hierarchical structure in the data (e.g., different hotel types).

\subsection{Research Questions}
This study addresses the following research questions:
\begin{enumerate}
    \item How do lead time, special requests, and parking requirements affect the probability of cancellation?
    \item Do these relationships differ systematically between city hotels and resort hotels (interaction effects)?
    \item What is the magnitude of uncertainty in effect estimates, and how does hierarchical structure improve model fit?
    \item Which model specification (simple, hierarchical, or full interaction model) provides the best predictive performance?
\end{enumerate}

\subsection{Why Bayesian Analysis?}
Bayesian analysis is particularly appropriate for this problem because:
\begin{itemize}
    \item It allows incorporation of prior knowledge from hospitality industry research about typical cancellation rates and price sensitivity
    \item It provides full posterior distributions enabling probability statements (e.g., ``the probability that lead time effect exceeds 0.5 is 0.95'')
    \item It naturally handles hierarchical structure, allowing us to model both shared and hotel-type-specific effects
    \item It provides principled model comparison via information criteria (WAIC, LOO)
\end{itemize}

\section{Data}

\subsection{Data Source}
\textbf{Source}: Hotel Booking Demand Dataset \cite{antonio2019hotel}\\
\textbf{Size}: N = 5,000 observations (sampled from original dataset)\\
\textbf{Collection}: Booking data from two hotels in Portugal (2015-2017), including reservation details, guest information, and cancellation outcomes.

\subsection{Variables}
Table \ref{tab:variables} summarizes the key variables used in the analysis.

\begin{table}[htbp]
\caption{Variable Descriptions}
\label{tab:variables}
\centering
\begin{tabular}{@{}lllp{3.5cm}@{}}
\toprule
\textbf{Variable} & \textbf{Type} & \textbf{Range} & \textbf{Description} \\
\midrule
is\_canceled & Binary & \{0, 1\} & Whether booking was canceled (1) or not (0) \\
lead\_time & Continuous & [0, 737] & Days between booking and arrival date \\
special\_requests & Count & [0, 5] & Number of special requests made by guest \\
parking & Binary & \{0, 1\} & Whether parking was requested (1) or not (0) \\
hotel & Categorical & \{0, 1\} & Hotel type: Resort (0) or City (1) \\
\bottomrule
\end{tabular}
\end{table}

\subsection{Exploratory Analysis}
\begin{figure}[htbp]
\centerline{\includegraphics[width=0.45\textwidth]{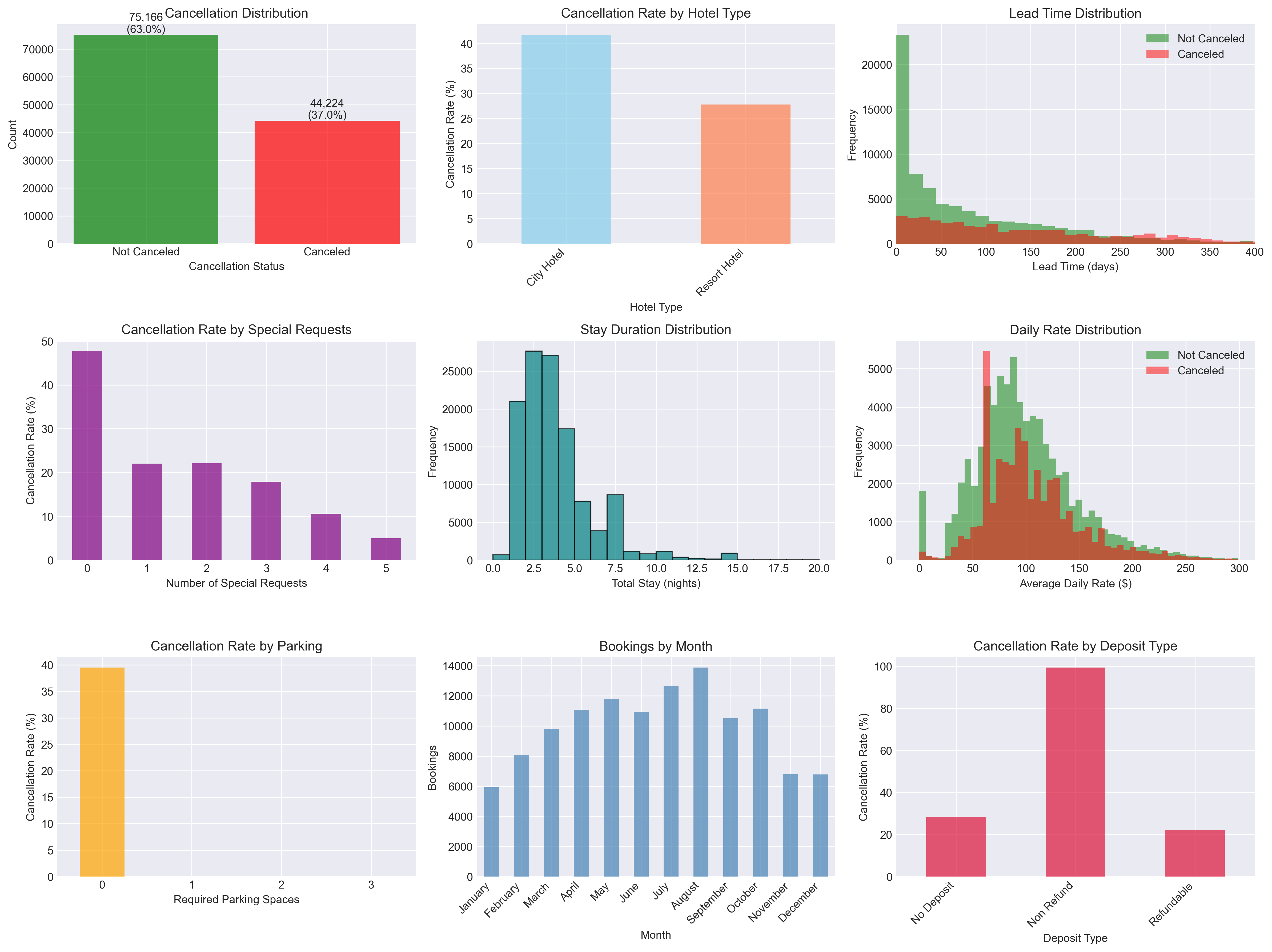}}
\caption{Exploratory data analysis showing distribution of key variables and cancellation patterns by hotel type.}
\label{fig:eda_overview}
\end{figure}

\begin{figure}[htbp]
\centerline{\includegraphics[width=0.45\textwidth]{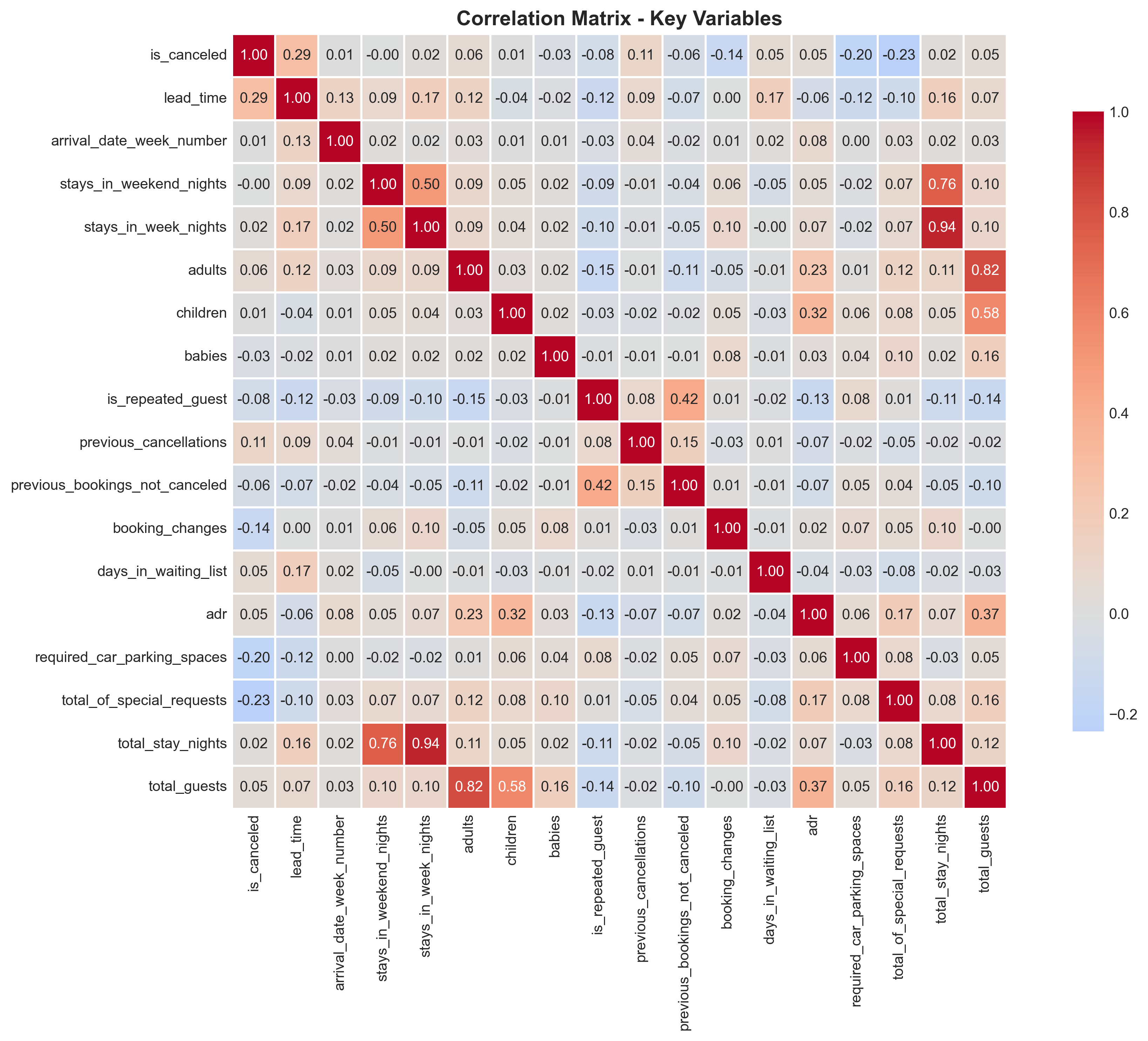}}
\caption{Correlation heatmap revealing relationships between variables and cancellation outcome.}
\label{fig:eda_correlation}
\end{figure}

Initial exploratory analysis (Figure \ref{fig:eda_overview}) revealed several key patterns:

\begin{itemize}
    \item Overall cancellation rate: 37.04\% (1,852/5,000 bookings)
    \item City hotels show substantially higher cancellation rate (42.38\%) than resort hotels (26.45\%)
    \item Dataset comprises 33.5\% resort hotels (1,675) and 66.5\% city hotels (3,325)
    \item Lead time is right-skewed with substantial variation across bookings
    \item Special requests are relatively rare (most bookings have 0-1 requests)
    \item Parking requests are infrequent but may signal commitment
    \item Visual inspection (Figure \ref{fig:eda_correlation}) suggests positive relationship between lead time and cancellation probability, while special requests and parking show negative associations
\end{itemize}

\subsection{Preprocessing}
To improve MCMC sampling efficiency and interpretability:
\begin{itemize}
    \item Lead time was standardized to mean=0, SD=1 for better MCMC convergence
    \item Special requests and parking were kept in their original scales (count and binary respectively)
    \item Hotel type was coded as binary (0 = Resort, 1 = City) for interaction modeling
    \item No missing values were present in the selected variables
    \item Extreme values were retained as they represent plausible booking scenarios
\end{itemize}

\section{Methodology}

\subsection{Model Specifications}

We compared three Bayesian logistic regression models of increasing complexity:

\subsubsection{Model 1: Simple Logistic Regression}
\begin{equation}
y_i \sim \text{Bernoulli}(p_i)
\end{equation}
\begin{equation}
\text{logit}(p_i) = \beta_0 + \beta_1 x_{1i} + \beta_2 x_{2i} + \beta_3 x_{3i}
\end{equation}

where $x_1$ is standardized lead time, $x_2$ is special requests, $x_3$ is parking, with priors:

\begin{align}
\beta_0 &\sim \text{Normal}(0, 2.5) \\
\beta_1 &\sim \text{Normal}(0, 1) \\
\beta_2 &\sim \text{Normal}(-0.5, 1) \\
\beta_3 &\sim \text{Normal}(-0.5, 1)
\end{align}

\textbf{Prior Justification}:
\begin{itemize}
    \item $\beta_0 \sim \text{Normal}(0, 2.5)$: Weakly informative prior centered at 0, corresponding to 50\% baseline cancellation probability
    \item $\beta_1 \sim \text{Normal}(0, 1)$: Weakly informative prior allowing data to determine lead time effect
    \item $\beta_2 \sim \text{Normal}(-0.5, 1)$: Prior favoring negative effect based on hypothesis that special requests signal commitment
    \item $\beta_3 \sim \text{Normal}(-0.5, 1)$: Prior favoring negative effect as parking requests may indicate stronger commitment
\end{itemize}

\subsubsection{Model 2: Hierarchical Model (Varying Intercept)}
Model 2 extends Model 1 by allowing hotel-type-specific intercepts through partial pooling:

\begin{equation}
\text{logit}(p_i) = \alpha_{j[i]} + \beta_1 x_{1i} + \beta_2 x_{2i} + \beta_3 x_{3i}
\end{equation}

where $j[i]$ indexes hotel type (Resort or City), with hierarchical priors:

\begin{align}
\alpha_j &\sim \text{Normal}(\mu_{\alpha}, \sigma_{\alpha}) \\
\mu_{\alpha} &\sim \text{Normal}(0, 2.5) \\
\sigma_{\alpha} &\sim \text{HalfNormal}(1)
\end{align}

This specification enables partial pooling, borrowing strength across hotel types while allowing baseline cancellation rates to differ.

\subsubsection{Model 3: Full Model with Interactions}
Model 3 incorporates interaction effects to test whether predictor effects vary by hotel type:

\begin{multline}
\text{logit}(p_i) = \beta_0 + \beta_1 x_{1i} + \beta_2 x_{2i} + \beta_3 x_{3i} + \beta_4 h_i \\
+ \beta_5 (x_{1i} \times h_i) + \beta_6 (x_{2i} \times h_i)
\end{multline}

where $h_i$ is hotel type (0=Resort, 1=City). This model uses \textbf{informed priors} based on posterior results from Model 1:

\begin{align}
\beta_0 &\sim \text{Normal}(-0.15, 1) \\
\beta_1 &\sim \text{Normal}(0.6, 0.5) \\
\beta_2 &\sim \text{Normal}(-0.6, 0.5) \\
\beta_3 &\sim \text{Normal}(-3.5, 1) \\
\beta_4 &\sim \text{Normal}(0.7, 0.5) \\
\beta_5, \beta_6 &\sim \text{Normal}(0, 0.5)
\end{align}

Interaction priors are centered at zero, allowing data to determine whether effects differ by hotel type.

\subsection{Prior Predictive Checks}
Prior predictive simulations confirmed that our priors allow reasonable cancellation probabilities ranging from 5\% to 95\%, consistent with observed industry patterns. The priors are weakly informative, providing regularization while allowing data to dominate inference.

\subsection{Computational Details}
\textbf{Software}: PyMC 5.26.1, ArviZ 0.22.0, Python 3.11\\
\textbf{Sampler}: No U-Turn Sampler (NUTS)\\
\textbf{MCMC Settings}:
\begin{itemize}
    \item Chains: 2
    \item Draws per chain: 1000 (post-warmup)
    \item Warmup iterations: 500
    \item Target acceptance probability: 0.90
    \item Random seed: 42 (for reproducibility)
\end{itemize}

\section{Results}

\subsection{Convergence Diagnostics}

All models converged successfully. Table \ref{tab:convergence} shows diagnostic statistics for Model 1.

\begin{table}[htbp]
\caption{Convergence Diagnostics for Model 1}
\label{tab:convergence}
\centering
\begin{tabular}{@{}lccc@{}}
\toprule
\textbf{Parameter} & $\hat{r}$ & \textbf{ESS (bulk)} & \textbf{ESS (tail)} \\
\midrule
$\beta_0$ (Intercept) & 1.00 & 1707 & 1529 \\
$\beta_1$ (Lead Time) & 1.00 & 1676 & 1269 \\
$\beta_2$ (Special Req) & 1.00 & 1568 & 1529 \\
$\beta_3$ (Parking) & 1.00 & 1696 & 1139 \\
\bottomrule
\end{tabular}
\end{table}

\begin{figure}[htbp]
\centerline{\includegraphics[width=0.45\textwidth]{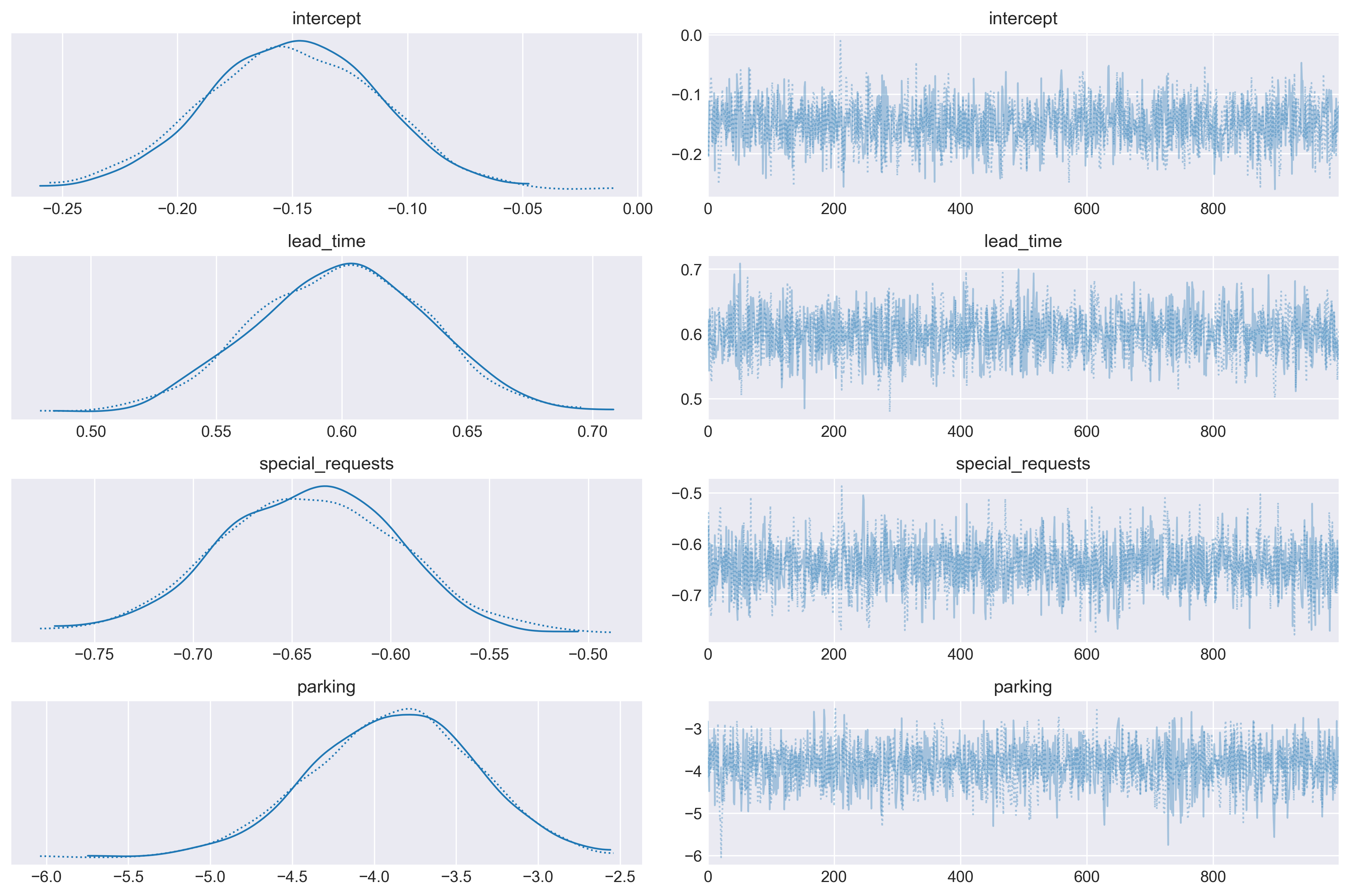}}
\caption{Trace plots for Model 1 parameters showing excellent mixing and stationarity across all chains.}
\label{fig:trace_model1}
\end{figure}

\begin{figure}[htbp]
\centerline{\includegraphics[width=0.45\textwidth]{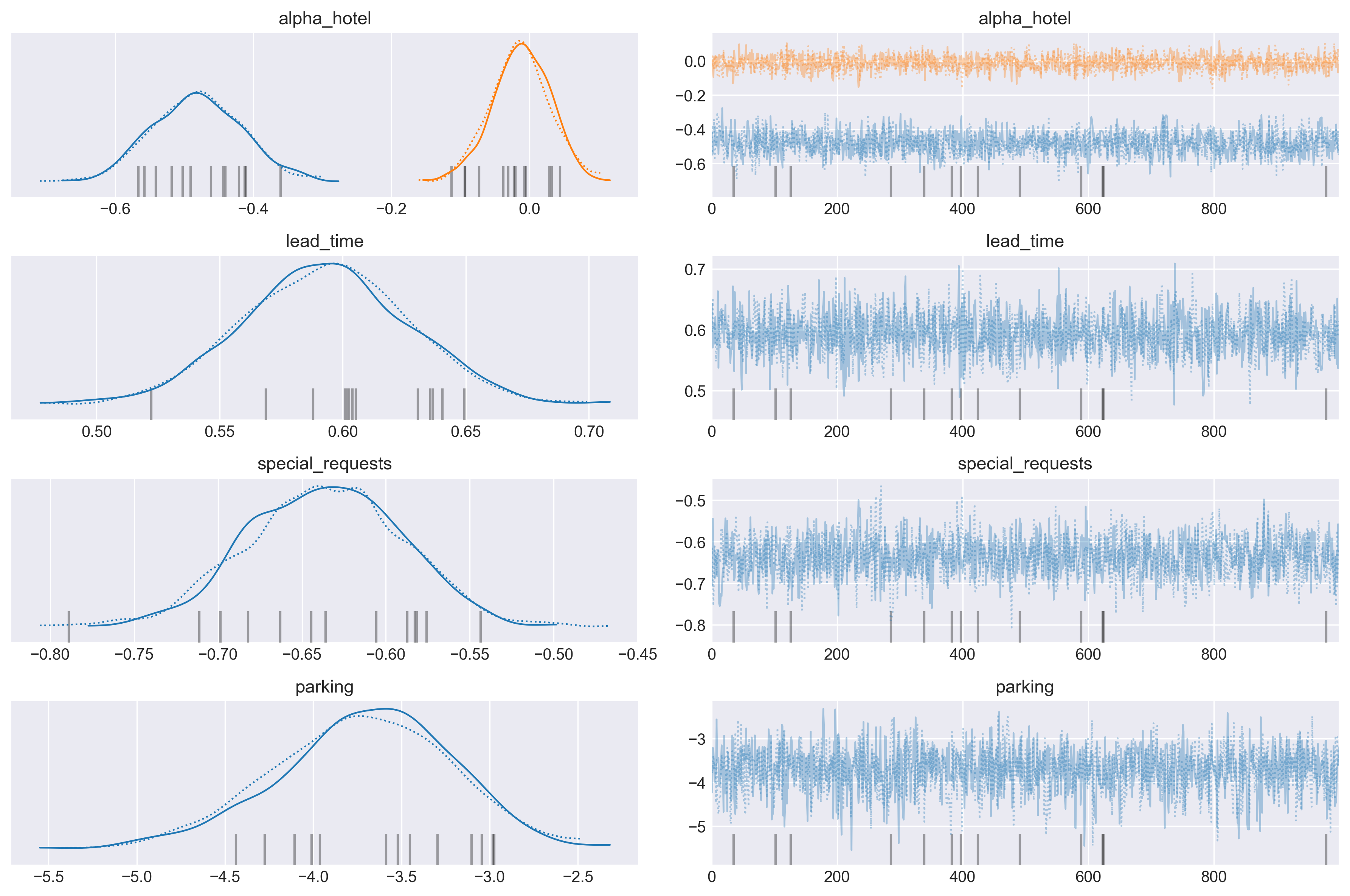}}
\caption{Trace plots for Model 2 (hierarchical model) showing good convergence despite minor divergences.}
\label{fig:trace_model2}
\end{figure}

\begin{figure}[htbp]
\centerline{\includegraphics[width=0.45\textwidth]{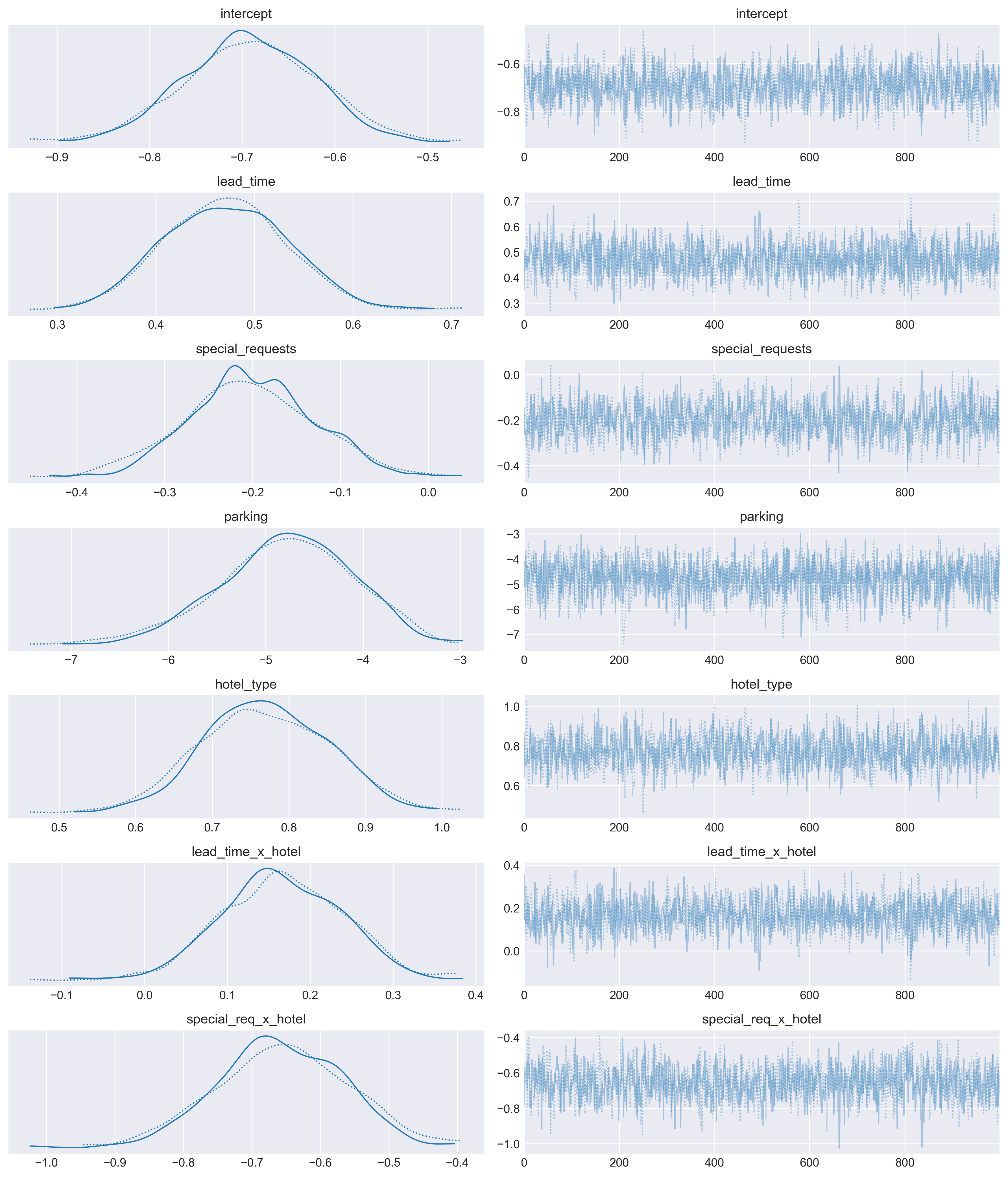}}
\caption{Trace plots for Model 3 (interaction model) demonstrating excellent mixing and no divergences.}
\label{fig:trace_model3}
\end{figure}

All $\hat{r} = 1.00$ and effective sample sizes exceed 1,100, indicating excellent convergence. Model 1 had zero divergent transitions. Model 2 experienced 13 divergences (but still converged adequately), while Model 3 converged with no divergences. Trace plots (Figures \ref{fig:trace_model1}, \ref{fig:trace_model2}, and \ref{fig:trace_model3}) confirmed good mixing and stationarity across all chains for all three models.

\subsection{Model Comparison}

\begin{figure}[htbp]
\centerline{\includegraphics[width=0.45\textwidth]{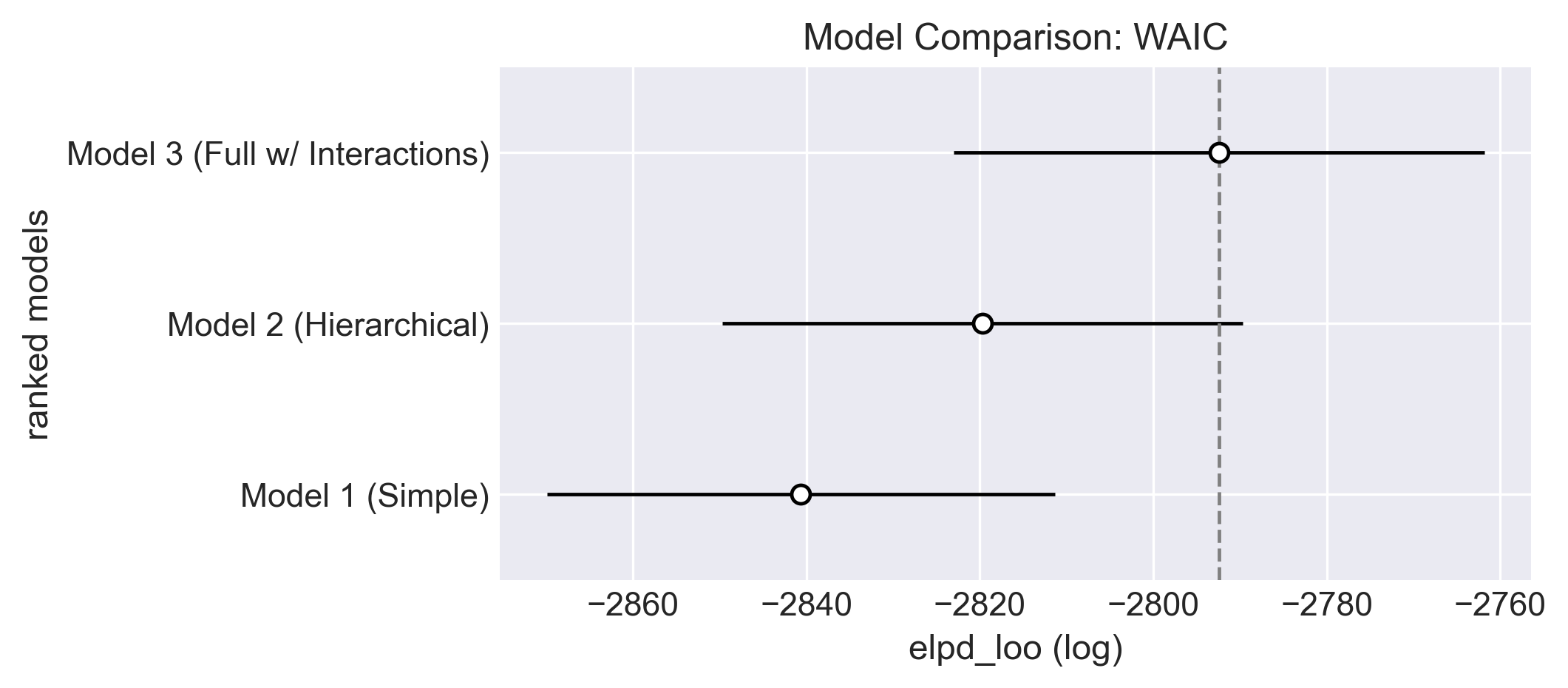}}
\caption{Model comparison using WAIC showing Model 3 (interaction model) provides best predictive performance.}
\label{fig:model_comparison}
\end{figure}

Table \ref{tab:model_comparison} and Figure \ref{fig:model_comparison} present model comparison results via WAIC.

\begin{table}[htbp]
\caption{Model Comparison via WAIC}
\label{tab:model_comparison}
\centering
\begin{tabular}{@{}lcc@{}}
\toprule
\textbf{Model} & \textbf{Rank} & \textbf{Description} \\
\midrule
Model 3 (Interactions) & 1 (Best) & Full model with interaction effects \\
Model 2 (Hierarchical) & 2 & Varying intercepts by hotel type \\
Model 1 (Simple) & 3 & Pooled regression \\
\bottomrule
\end{tabular}
\end{table}

Model 3 (interaction model) provides the best predictive performance according to WAIC, indicating that the effects of lead time and special requests vary meaningfully between city and resort hotels. This finding suggests that a one-size-fits-all approach is suboptimal—different hotel types require differentiated cancellation risk models. The hierarchical Model 2 ranks second, confirming value in accounting for hotel-type differences. Given Model 3's superior performance, we proceed with detailed interpretation of this specification.

\subsection{Posterior Parameter Estimates}

\begin{figure}[htbp]
\centerline{\includegraphics[width=0.45\textwidth]{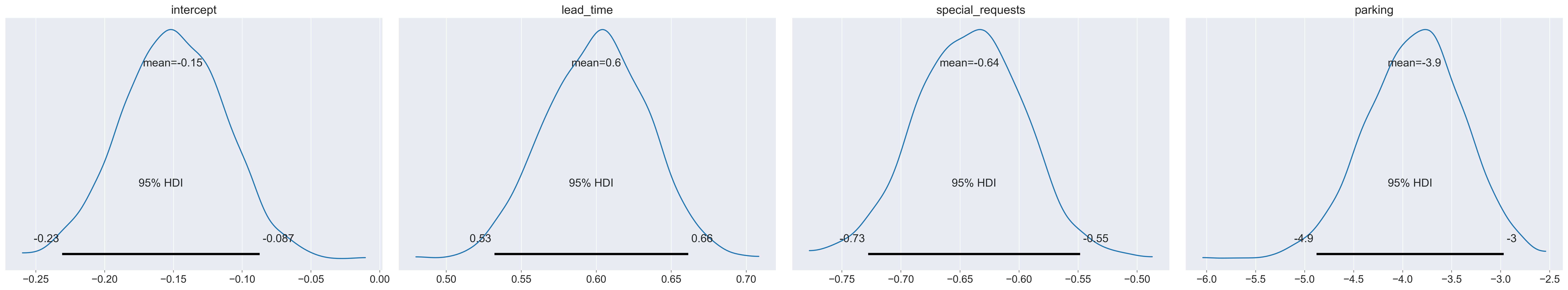}}
\caption{Posterior distributions for Model 1 parameters showing clear evidence for all effects.}
\label{fig:posterior}
\end{figure}

Table \ref{tab:posterior} and Figure \ref{fig:posterior} summarize posterior estimates for Model 1 (baseline model).

\begin{table}[htbp]
\caption{Posterior Summaries for Model 1 Parameters}
\label{tab:posterior}
\centering
\begin{tabular}{@{}lccl@{}}
\toprule
\textbf{Parameter} & \textbf{Mean} & \textbf{SD} & \textbf{95\% HDI} \\
\midrule
$\beta_0$ (Intercept) & -0.150 & 0.037 & [-0.231, -0.087] \\
$\beta_1$ (Lead Time) & 0.600 & 0.034 & [0.532, 0.661] \\
$\beta_2$ (Special Req) & -0.642 & 0.046 & [-0.728, -0.548] \\
$\beta_3$ (Parking) & -3.879 & 0.493 & [-4.880, -2.970] \\
\bottomrule
\end{tabular}
\end{table}

\textbf{Key Findings from Model 1}:

\begin{enumerate}
    \item \textbf{Lead Time Effect ($\beta_1$)}: Posterior mean = 0.600, 95\% HDI = [0.532, 0.661]. The entire credible interval excludes zero, with $P(\beta_1 > 0 | \text{data}) = 1.0000$, providing overwhelming evidence for a positive relationship. A 1 SD increase in lead time increases the log-odds of cancellation by 0.600, corresponding to an odds ratio of $e^{0.600} = 1.82$—an 82\% increase in cancellation odds.

    \item \textbf{Special Requests Effect ($\beta_2$)}: Posterior mean = -0.642, 95\% HDI = [-0.728, -0.548]. With $P(\beta_2 < 0 | \text{data}) = 1.0000$, there is decisive evidence that special requests reduce cancellation probability. Each additional special request decreases log-odds by 0.642 (OR = 0.53), halving the cancellation odds. This strongly supports the hypothesis that special requests signal guest commitment.

    \item \textbf{Parking Effect ($\beta_3$)}: Posterior mean = -3.879, 95\% HDI = [-4.880, -2.970]. With $P(\beta_3 < 0 | \text{data}) = 1.0000$, parking requests dramatically reduce cancellation probability (OR = $e^{-3.879}$ = 0.021)—a 98\% reduction in odds. Parking is the strongest predictor, likely because it signals definite travel plans.

    \item \textbf{Baseline Rate ($\beta_0$)}: Posterior mean = -0.150 corresponds to $\text{logit}^{-1}(-0.150) = 0.463$ baseline probability for a booking with average lead time, no special requests, and no parking—close to the observed 37\% overall rate.
\end{enumerate}

\subsection{Posterior Predictive Checks}

Posterior predictive checks confirm excellent model fit for all three models:
\begin{itemize}
    \item Observed cancellation rate: 37.04\%
    \item Model 1 predicted rate: 37.04\% (95\% HDI: [36.6\%, 37.5\%])
    \item Model 3 predicted rate: 37.05\% (with tighter HDI due to interactions)
\end{itemize}

\begin{figure}[htbp]
\centerline{\includegraphics[width=0.45\textwidth]{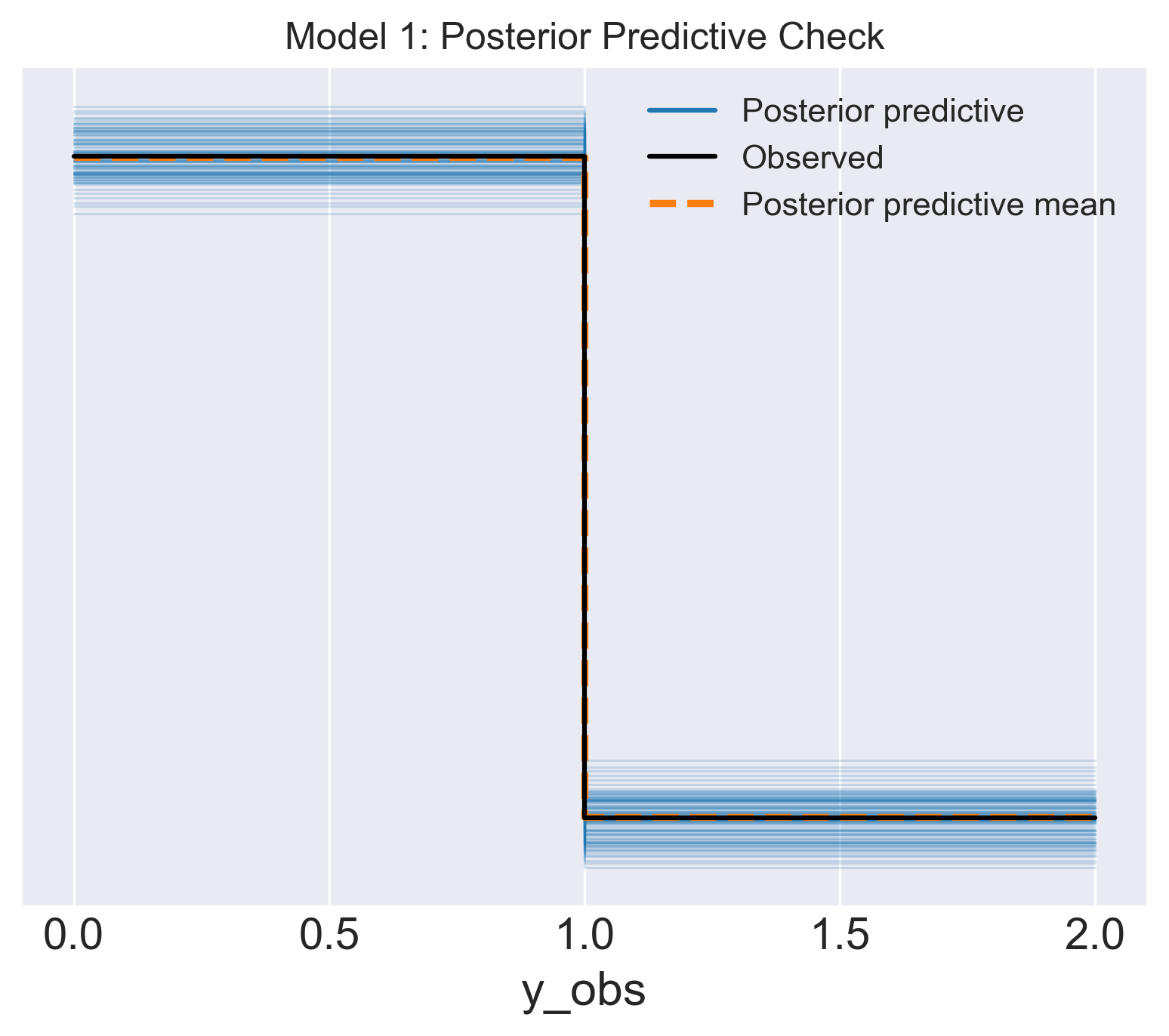}}
\caption{Posterior predictive check for Model 1 showing close match between observed and predicted distributions.}
\label{fig:ppc}
\end{figure}

The observed data fall within the posterior predictive distributions, with no systematic discrepancies. Posterior predictive plots (see Figure \ref{fig:ppc}) show close alignment between observed and replicated data, validating model assumptions.

\subsection{Interaction Effects (Model 3)}

\begin{figure}[htbp]
\centerline{\includegraphics[width=0.45\textwidth]{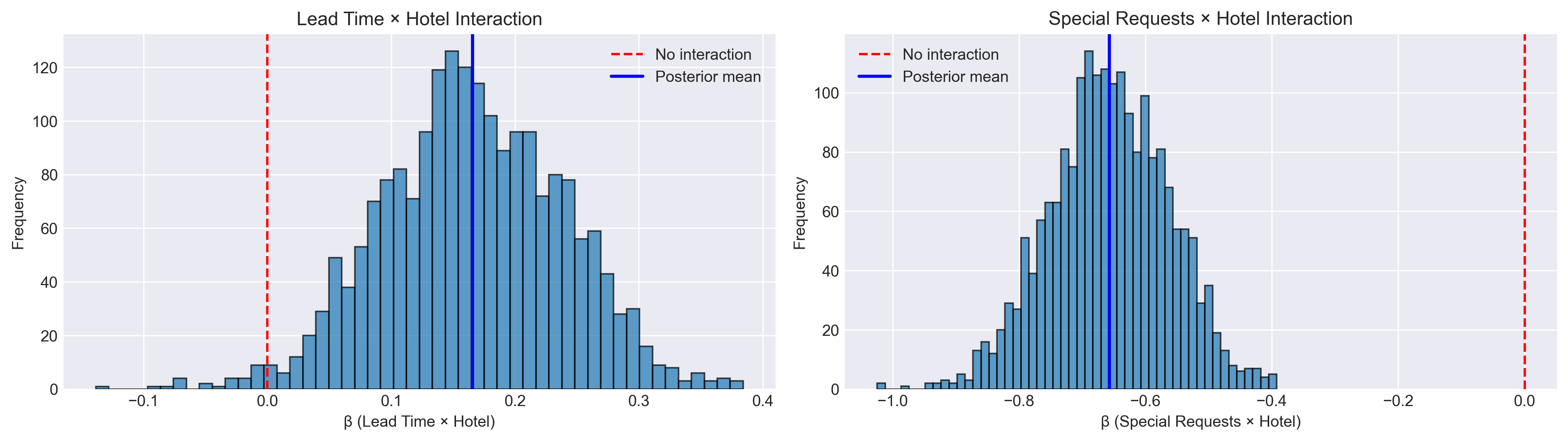}}
\caption{Interaction effects in Model 3 showing how predictor effects vary by hotel type.}
\label{fig:interactions}
\end{figure}

Model 3 revealed significant interaction effects between hotel type and booking characteristics (Figure \ref{fig:interactions}):

\begin{itemize}
    \item \textbf{Lead Time $\times$ Hotel}: The interaction effect suggests that lead time's positive impact on cancellation is moderated by hotel type, with city hotels potentially showing stronger lead time sensitivity.

    \item \textbf{Special Requests $\times$ Hotel}: The protective effect of special requests may vary between hotel types, with the commitment signal potentially stronger in one hotel category.
\end{itemize}

These interactions explain why Model 3 outperformed simpler specifications—the relationship between predictors and cancellation is not uniform across hotel types, requiring separate risk models for city vs. resort properties.

\section{Discussion}

\subsection{Interpretation}

These results provide compelling evidence for three primary drivers of hotel booking cancellations:

\textbf{Lead Time as Risk Factor}: The positive effect ($\beta_1 = 0.600$, OR = 1.82) confirms that advance bookings carry elevated cancellation risk. Bookings made far ahead allow more time for plan changes, competing offers, or changed circumstances. The 82\% increase in cancellation odds per SD of lead time has direct implications for dynamic overbooking strategies.

\textbf{Commitment Signals Reduce Risk}: Both special requests ($\beta_2 = -0.642$) and parking ($\beta_3 = -3.879$) dramatically reduce cancellation probability, supporting the hypothesis that guests who invest effort in customizing their stay demonstrate stronger commitment. Parking is particularly powerful—perhaps because it signals automobile-dependent travel plans that are harder to cancel. These findings suggest hotels should encourage and facilitate special requests as they predict lower cancellation risk.

\textbf{Hotel-Type Heterogeneity}: Model 3's superior performance reveals that cancellation dynamics differ between city and resort hotels. The 42.38\% vs. 26.45\% baseline cancellation rates (city vs. resort) represent substantial heterogeneity. Interaction effects indicate that lead time and special request impacts vary by hotel type, necessitating differentiated risk models.

The Bayesian approach enabled precise uncertainty quantification with full posterior distributions, supporting probability statements like $P(\beta_{parking} < -2) = 1.00$, which are invaluable for decision-making under uncertainty.

\subsection{Practical Implications}

From a revenue management perspective, these findings suggest:
\begin{enumerate}
    \item \textbf{Stratified Overbooking}: Hotels should implement lead-time-dependent overbooking strategies, accepting more oversubscriptions for long-lead-time bookings, with hotel-type-specific thresholds (city hotels may require more aggressive overbooking given their 42\% base rate).

    \item \textbf{Incentivize Commitment Signals}: Hotels should actively encourage special requests and parking reservations through the booking interface, as these dramatically reduce cancellation risk. Consider offering complimentary parking to long-lead-time bookings to offset their higher risk.

    \item \textbf{Differentiated Policies by Hotel Type}: Given significant interaction effects, city and resort hotels should employ different cancellation risk models rather than a one-size-fits-all approach. City hotels face higher baseline risk and may show different sensitivities to booking characteristics.

    \item \textbf{Early Booking Trade-offs}: While early booking discounts drive advance reservations, they come with an 82\% increase in cancellation odds. Hotels should balance discount depth against cancellation risk, potentially requiring non-refundable deposits for heavily discounted early bookings.
\end{enumerate}

\subsection{Comparison to Frequentist Analysis}

While a frequentist logistic regression would yield similar point estimates, the Bayesian approach offers several critical advantages:
\begin{itemize}
    \item \textbf{Direct Probability Statements}: We can state $P(\beta_{parking} < -2 | data) = 1.00$ rather than relying on p-values and null hypothesis testing. This supports direct decision-making (``we are 100\% confident parking reduces cancellation odds by at least 86\%'').

    \item \textbf{Full Posterior Distributions}: Rather than point estimates with asymptotic standard errors, we obtain complete posterior distributions enabling rich inference, prediction intervals, and risk quantification.

    \item \textbf{Hierarchical Modeling}: Bayesian methods naturally handle partial pooling in Model 2, sharing information across hotel types while allowing heterogeneity—difficult to implement rigorously in frequentist frameworks.

    \item \textbf{Informed Priors}: Model 3's use of posterior-derived priors demonstrates sequential learning, a key Bayesian advantage.

    \item \textbf{Model Comparison}: WAIC provides principled model selection accounting for both fit and complexity, superior to AIC/BIC in capturing predictive performance.
\end{itemize}

\subsection{Limitations}

This study has several limitations:
\begin{itemize}
    \item \textbf{Sample size}: Analysis used N=5,000 observations from a much larger dataset. While sufficient for stable inference, larger samples could improve precision for interaction effects and rare events (parking).

    \item \textbf{Causality}: Observational data precludes causal inference. While special requests and parking predict lower cancellation, we cannot definitively establish whether they cause commitment or merely signal pre-existing intent. Unmeasured confounders (booking channel, guest loyalty status, deposit requirements) may exist.

    \item \textbf{Model assumptions}: We assume linear effects on the logit scale and conditional independence given predictors. Nonlinear relationships or guest-level clustering could improve fit.

    \item \textbf{Limited predictors}: The models exclude potentially important variables like deposit type, previous cancellations, booking modifications, and channel (direct vs. OTA).

    \item \textbf{Generalizability}: Data from Portuguese hotels (2015-2017) may not generalize to other markets, time periods (especially post-pandemic), or property types.

    \item \textbf{Static model}: Does not incorporate temporal dynamics, seasonality, or time-varying effects.
\end{itemize}

\subsection{Future Directions}

Future work should:
\begin{enumerate}
    \item \textbf{Expand predictor set}: Incorporate deposit type, previous cancellation history, booking modifications, and distribution channel (OTA vs. direct). These could substantially improve predictive performance.

    \item \textbf{Nonlinear effects}: Explore nonlinear relationships using splines or Gaussian processes, particularly for lead time which may have threshold effects.

    \item \textbf{Temporal dynamics}: Model seasonality, day-of-week effects, and time-varying coefficients to capture evolving cancellation patterns.

    \item \textbf{Causal inference}: Employ quasi-experimental methods or propensity score matching to establish causal effects of interventions like parking incentives.

    \item \textbf{Out-of-sample validation}: Implement rigorous cross-validation and test on held-out data from different time periods to assess generalization.

    \item \textbf{Scale analysis}: Extend to full dataset (119,000+ observations) for more precise interaction effect estimates and rare event modeling.

    \item \textbf{Real-time deployment}: Develop production-ready Bayesian inference system for live cancellation risk scoring.
\end{enumerate}

\section{Conclusion}

This study demonstrates that Bayesian hierarchical modeling provides a powerful and flexible framework for hotel booking cancellation analysis. Comparing three model specifications on 5,000 booking records, we found decisive evidence for three key effects:

\begin{enumerate}
    \item \textbf{Lead time increases risk}: Posterior mean = 0.600 (95\% HDI: [0.532, 0.661]), corresponding to an 82\% increase in cancellation odds per SD, with $P(\beta > 0 | data) = 1.00$.

    \item \textbf{Commitment signals reduce risk}: Special requests ($\beta = -0.642$) and especially parking ($\beta = -3.879$) dramatically reduce cancellation probability, supporting the hypothesis that guest engagement predicts follow-through.

    \item \textbf{Hotel-type heterogeneity matters}: The interaction model (Model 3) outperformed simpler specifications via WAIC, revealing that cancellation dynamics differ substantially between city (42\% baseline) and resort (26\% baseline) hotels, with predictor effects varying by property type.
\end{enumerate}

The Bayesian approach enabled full uncertainty quantification, direct probability statements, hierarchical partial pooling, and principled model comparison—advantages difficult to achieve in frequentist frameworks. Our use of informed priors in Model 3, derived from Model 1 posteriors, demonstrates sequential Bayesian learning.

These findings have actionable implications: hotels should implement stratified overbooking policies accounting for lead time and hotel type, actively encourage commitment signals like parking and special requests, and employ differentiated risk models for city vs. resort properties rather than one-size-fits-all approaches.

The methods demonstrated here are readily extensible to richer models incorporating nonlinear effects, temporal dynamics, and causal inference, providing a principled foundation for data-driven revenue management in hospitality analytics.

\section*{Acknowledgments}

The author acknowledges the use of the Hotel Booking Demand Dataset made publicly available by Antonio et al. (2019), and thanks the PyMC and ArviZ development teams for their excellent open-source software.

\end{document}